\documentclass[10pt,conference]{IEEEtran}
\IEEEoverridecommandlockouts

\usepackage[T1]{fontenc}
\usepackage{xspace}
\usepackage{xcolor}
\usepackage{amsfonts}
\usepackage{graphicx}
\usepackage{amsmath,amssymb}
\usepackage{hyperref}
\usepackage{booktabs}
\usepackage{cleveref}
\usepackage{xurl}
\usepackage[inline]{enumitem}
\usepackage{tikz}
\usepackage{soul}
\usepackage{xcolor}
\usepackage{xcolor,colortbl}
\usetikzlibrary{shapes.geometric,backgrounds,calc,positioning,fit,patterns}

\makeatletter
\newcommand{\linebreakand}{%
  \end{@IEEEauthorhalign}
  \hfill\mbox{}\par
  \mbox{}\hfill\begin{@IEEEauthorhalign}
}
\makeatother

\newcommand{\numfaults}{13\xspace}

\usepackage{comment}
\usepackage{ifthen}

\newboolean{showcomments}
\setboolean{showcomments}{true}
\ifthenelse{\boolean{showcomments}}
{\newcommand{\nb}[2]{
  \fcolorbox{black}{yellow}{\bfseries\sffamily\scriptsize#1}
  {\sf\small$\blacktriangleright$\textit{#2}$\blacktriangleleft$}
 }
 
}
{\newcommand{\nb}[2]{}
 
}
\usepackage{subcaption}

\newcommand\Simulink{Simulink\textsuperscript{\tiny\textregistered}\xspace}
\newcommand\SimulinkFaultAnalyzer{Simulink\textsuperscript{\tiny\textregistered} Fault Analyzer\textsuperscript{\tiny TM}\xspace}
\newcommand\Simscape{Simscape\textsuperscript{\tiny TM}\xspace}

\usepackage[framemethod=TikZ]{mdframed}
\mdfsetup{nobreak=false}
\newenvironment{Answer}[1][]{%
  \ifstrempty{#1}%
  {\mdfsetup{%
    frametitle={%
      \tikz[baseline=(current bounding box.east),outer sep=0pt]
      \node[line width=0pt,anchor=east,rectangle,draw=white,fill=white]
    ;}}
  }%
  {\mdfsetup{%
    frametitle={%
      \tikz[baseline=(current bounding box.east),outer sep=0pt]
      \node[anchor=east,rectangle,draw=white,fill=white]
    {\strut #1};}}%
  }%
  \mdfsetup{innertopmargin=-5pt,linecolor=black,%
            linewidth=0.5pt,topline=true,%
            frametitleaboveskip=\dimexpr-\ht\strutbox\relax,skipabove=.5\topskip,skipbelow=.2\topskip}
  \begin{mdframed}[nobreak=false]\relax
  }{\end{mdframed}}

\usepackage{multirow}
\begin{document}

\title{Failure Modes and Effects Analysis:\\ An Experience from the E-Bike Domain}

\author{\IEEEauthorblockN{Andrea Bombarda\textsuperscript{\textsection}}
\IEEEauthorblockA{\textit{University of Bergamo}\\
Bergamo, Italy \\
andrea.bombarda@unibg.it}
\and
\IEEEauthorblockN{Federico Conti\textsuperscript{\textsection}}
\IEEEauthorblockA{\textit{University of Bergamo}\\
Bergamo, Italy \\
f.conti12@studenti.unibg.it}
\and
\IEEEauthorblockN{Marcello Minervini}
\IEEEauthorblockA{\textit{University of Bergamo}\\
Bergamo, Italy \\
marcello.minervini@unibg.it}\linebreakand
\IEEEauthorblockN{Aurora Zanenga}
\IEEEauthorblockA{\textit{University of Bergamo}\\
Bergamo, Italy \\
a.zanenga@studenti.unibg.it}
\and
\IEEEauthorblockN{Claudio Menghi}
\IEEEauthorblockA{\textit{University of Bergamo}\\
Bergamo, Italy \\
\textit{McMaster University}\\
Hamilton, ON, Canada \\
claudio.menghi@unibg.it}
}

\maketitle              
\begingroup\renewcommand\thefootnote{\textsection}
\footnotetext{These authors contributed equally}
\endgroup
\thispagestyle{plain}
\pagestyle{plain}

\begin{abstract}
Software failures can have catastrophic and costly consequences.
Functional Failure Mode and Effects Analysis (FMEA) is a standard technique used within Cyber-Physical Systems (CPS) to identify software failures and assess their consequences.
Simulation-driven approaches have recently been shown to be effective in supporting FMEA.
However, industries need evidence of the effectiveness of these approaches to increase practical adoption. 
This industrial paper presents our experience with using FMEA to analyze the safety of a CPS from the e-Bike domain.
We used Simulink Fault Analyzer, an
industrial tool that supports engineers with FMEA.
We identified 13 realistic faults, modeled them, and analyzed their effects.
We sought expert feedback to analyze the appropriateness of our models and the effectiveness of the faults in detecting safety breaches.
Our results reveal that for the faults we identified, our models were accurate or contained minor imprecision that we subsequently corrected.
They also confirm that FMEA helps engineers improve their models. 
Specifically, the output provided by the simulation-driven support for 38.4\% (5 out of 13) of the faults did not match the engineers' expectations, helping them discover unexpected effects of the faults.
We present a thorough discussion of our results and ten lessons learned.
Our findings are useful for software engineers who work as \Simulink engineers,  use the \SimulinkFaultAnalyzer, or work as safety analysts.
\end{abstract}


\section{Introduction}
\label{sec:introduction}

Failure Mode and Effects Analysis (FMEA) is widely used in Cyber-Physical Systems (CPS) development to analyze the safety of software products~\cite{reifer1979software,valyayev2024fmea,fmeasuccess}.
For example, it has been used in several NASA programs~\cite{office1966procedure,mode1993effects} such as Apollo, Viking, Voyager, Magellan, Galileo, and Skylab and it is also used by Ford~\cite{ford2003ford}.
FMEA requires identifying potential failure modes (Failure Mode - FM) and analyzing their causes and effects (Effects Analysis - EA).

As CPSs become more safety-critical, safety assurance and safety analysis are increasingly becoming essential in software engineering~\cite{Chechik2022,Rodriguez2023,Zhu2023}.
To support software safety analysis from the early stages of the software development process, researchers have developed approaches that embed safety analysis within system modeling~\cite{joshi2006model,lisagor2011model,Goddard2000,Rebello2010,Krishnan2020,Bartocci2022}. 
For example, Simulation-driven FMEA~\cite{rhein2024simulation} is an automated software engineering solution that checks the effects of the faults defined in an FMEA table by performing simulations of models enriched with faults.
The results of these simulations enable engineers to analyze the effects of the faults and, depending on their criticality, develop strategies that mitigate them.
Simulation-driven FMEA provided encouraging results on a pivotal study that represented a flight control system of an unmanned ultralight helicopter~\cite{rhein2024simulation}; it is currently integrated within the \SimulinkFaultAnalyzer~\cite{SimulinkFaultAnalyzer}. 
The \SimulinkFaultAnalyzer is a tool for the safety management of software systems developed in \Simulink; it enables the connection of faults, hazards, and mitigation logic.

Safety analysis experts need empirical evidence regarding the benefits and limitations of the different techniques to decide whether to incorporate them into their working pipeline.
Simulation-driven FMEA~\cite{rhein2024simulation} and the \SimulinkFaultAnalyzer have been recently introduced within \Simulink (R2023b). 
To assess the industrial applicability of Simulation-driven FMEA, it is paramount to empirically evaluate its effectiveness and provide practitioners with guidelines and lessons learned~\cite{Shaukat2010}, as widely recognized by the research and industrial software engineering communities~\cite{Shaukat2010,dyba2005evidence,kitchenham2004evidence,panichella2018large,sayyad2013parameter,melo2019empirical,7107470,8718592,valle2025industrial}. 
This need is of primary importance for \Simulink~\cite{boll2021characteristics,shrestha2023replicability,boll2020replicability,boll2024replicability}, as Simulink projects and models are typically industrial and usually not shared or publicly available due to corporate policies~\cite{badreddin2013modeling,ding2014open}.

\begin{figure*}[t]
    \centering   \includegraphics[width=.98\textwidth]{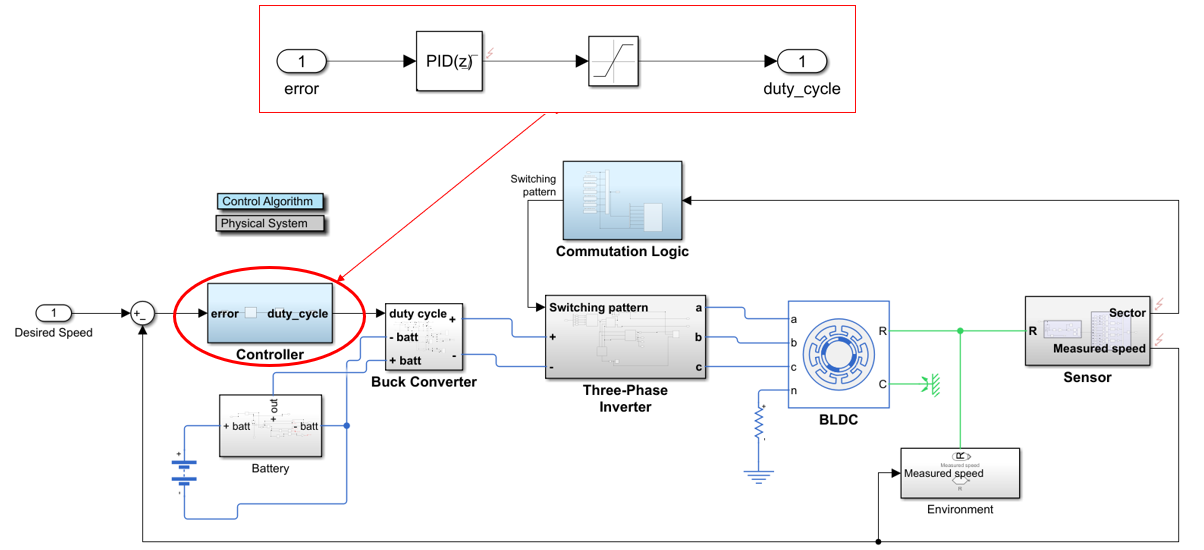}
    \caption{\Simulink model for the e-Bike.}
    \label{fig:pwm}
\end{figure*}

To mitigate this need, this industry track paper reports on the usefulness of Simulation-driven FMEA on a study subject from the e-Bikes domain. 
Specifically, our study subject is a \Simulink model related to the software controller of an e-Bike. 
It is developed in the EU-funded project MOST~\cite{MOST}, that aims at improving ``green'' mobility solutions, including e-Bikes, and involving several companies, such as Brembo~\cite{Brembo} and Pirelli~\cite{Pirelli}.
E-Bikes are vehicles that support riders with power from an electric motor. 
Our case study is relevant because 
(a)~the e-Bike software is often designed in \Simulink~\cite{E-Bosch};
(b)~e-Bikes have a considerable market size (27.15 USD billion in 2022, expected to grow to USD 82.84 billion by 2030~\cite{E-BikeMarket});
(c)~they are software-intensive systems~\cite{SofwareBike}, as the software controls the electrical motor of the e-Bike and regulates its speed and possibly the regenerative power;
(d)~they are safety-critical systems which must comply with safety regulations and requirements~\cite{Schepers2018safety,Schleinitz2017}; and
(e)~our results and lessons learned are based on the interaction with an expert (the third author of this paper) who is designing the software for the controller of an e-Bike within the context of the MOST project~\cite{MOST} that involves several industrial partners who specialize in the e-Bike domain as well as on discussions with \Simulink engineers.
Our findings are useful for software engineers who work as \Simulink engineers, use the \SimulinkFaultAnalyzer, or work as safety analysts.

To evaluate the usefulness of Simulation-driven FMEA, we considered 13 realistic faults; we (a)~ modeled them within the \Simulink model using the \SimulinkFaultAnalyzer and (b)~analyzed their effects.
This is a significant number of software faults considering the application domain.
Developing safety-critical CPS is costly and time-consuming~\cite {EBikeSoftware}.
To design these faults, we needed a holistic view of the system, including its motor and battery.
Acquiring this knowledge is time-consuming.
It required us to interact with our domain expert and thoroughly study the domain to obtain the domain knowledge to operate in this field.
We retrospectively estimate this effort as six months of work.

To assess the benefits of Simulation-driven FMEA, we interviewed our expert to ensure that the models of our faults reflected their intended behavior.
Our results reveal that all faults were modeled correctly (53.8\%) or with minor imprecision (46.2\%) that were fixed.
Our discussion with \Simulink experts enabled us to fix the models for these faults.
We interviewed the expert to understand the expected consequences of the faults and compare them with those provided by Simulation-driven FMEA.
For over 38\% of the faults, the output provided by the simulation-driven support for FMEA provided by the \SimulinkFaultAnalyzer did not match the engineer's expectation, thereby leading to the field engineers discovering unexpected effects of the faults.
To provide a thorough interpretation of our results, we discussed them with  \Simulink engineers, where we presented the objective of the work, our results, and collected their informal feedback and interpretation of the results.
These discussions led to the formalization of ten lessons learned --- i.e., five \Simulink specific (useful for \Simulink engineers and users of the \SimulinkFaultAnalyzer) and five \Simulink agnostic (useful for safety analysts).

The remainder of this paper is organized as follows.
\Cref{sec:casestudy} introduces our e-Bike study subject.
\Cref{sec:sdfmea} summarizes Simulation-driven FMEA.
\Cref{sec:evaluation} presents our empirical evaluation of Simulation-driven FMEA.
\Cref{sec:discussion} discusses our results and presents lessons learned.
\Cref{sec:related} presents related work.
\Cref{sec:conclusions} concludes our work.

\section{The E-Bike Study Subject}
\label{sec:casestudy}

\Cref{fig:pwm} presents our study subject: The \Simulink model of an e-Bike~\cite{Marci}.
This model is developed within the context of a project~\cite{MOST} that involves several companies (e.g., Brembo~\cite{Brembo} and Pirelli~\cite{Pirelli}) to improve ``green'' mobility solutions.
Its development time is approximately 100 hours~\cite{Corti2024}.

The controlled system consists of several components:
\begin{itemize}
    \item The \emph{Environment} component models external forces (e.g., the friction and aerodynamic drag). It models external loads, mimicking the actual resistance an e-Bike would encounter during operation and that would influence motor performance. 
    Its input is the speed of the e-Bike (\emph{Measured speed}).
    Its output is a signal that simulates the effects of friction and aerodynamic torque (\emph{R}).
    \item The \emph{Brushless Direct Current Motor (BLDC)} component converts electrical energy into rotational motion and torque.
    Its inputs are the currents applied to the three phases of the BLDC ($a$, $b$, $c$) and the neutral point ($n$). 
    Its outputs are the mechanical rotational conserving torque associated with the motor rotor (\emph{R}) and motor case (\emph{C}).
    \item The \emph{Sensor} component monitors the status of the e-Bike by measuring its torque, and the speed and sector motor's rotor.
    Its input is the torque of the rotor (\emph{R}).
    Its outputs are the active sector (\emph{Sector}) of the BLDC motor and the e-Bike speed (\emph{Measured speed}).
    \item The \emph{Battery} component stores and retrieves electrical energy. 
    It receives as input the current that recharges the battery (\emph{-~Batt}).
    It outputs a current from the energy stored within the battery (\emph{+~Batt}) to feed the motor. 
    \item The \emph{Three-Phase Inverter} component converts the direct current into alternating current.
    It receives two inputs(\emph{-~Batt} and \emph{+~Batt}) from the buck converter. It acts on the motor, depending on the direct current (DC) signal (\emph{Switching pattern}) received from the software controller (\emph{Controller}).
    It outputs the currents applied to the three phases of the BLDC ($a$, $b$, $c$).
    
    \item The \emph{Commutation Logic} component determines the voltage to be supplied by the three-phase inverter that is used to regulate the stator winding currents and to create a rotating magnetic field that drives the rotor of the BLDC motor.
    Its input is the active sector of the BLDC (\emph{Sector}).
    Its outputs are the control signals to be sent to the electronic switches into the three-phase inverter (\emph{Switching pattern}).
    \item The \emph{Buck Converter} component converts a higher input voltage to a lower output voltage based on the duty cycle value.
    Its inputs are the duty cycle (\emph{duty\_cycle}) and the voltage from the battery (\emph{- Batt}, \emph{+ Batt}).
    Its output is the voltage to be fed into the Three-Phase Inverter (\emph{-}, \emph{+}).
    \item The \emph{(Software) Controller} component regulates the behavior of the BLDC motor to ensure that the measured speed (the model's output) matches the desired speed selected by the user (the model's input).
    Its input is the error difference (\emph{Error}) between the desired speed and the measured speed.
    Its output is the duty cycles (\emph{duty\_cycle}). 
\end{itemize}

Engineers \emph{did not} analyze the \emph{safety} of their system design.
This work aims to analyze how Simulation-driven FMEA and the tool support provided by \SimulinkFaultAnalyzer can enable engineers to develop a safer system design.

\section{Simulation-driven FMEA}
\label{sec:sdfmea}
\SimulinkFaultAnalyzer enables engineers to perform FMEA safety analysis by providing support for (a)~defining potential failure modes (FM) by modeling faults (\Cref{sec:modeling}) and (b)~effect analysis (EA) by simulating their consequences (\Cref{sec:effects}).

\subsection{Support for Failure Mode - FM}
\label{sec:modeling}
\SimulinkFaultAnalyzer enables engineers to extend the design model with faults.
For example, \Cref{fig:pwm} presents three faults that are inserted on the \emph{Sector}, \emph{Measured speed}, and \emph{PID(z)} signals.
\SimulinkFaultAnalyzer graphically identifies faults with a lightning symbol on the signal on which the fault is applied (or on a component, for the \Simscape~\cite{simscape} toolbox).

\begin{figure}[t]
    \centering   \includegraphics[width=.98\columnwidth]{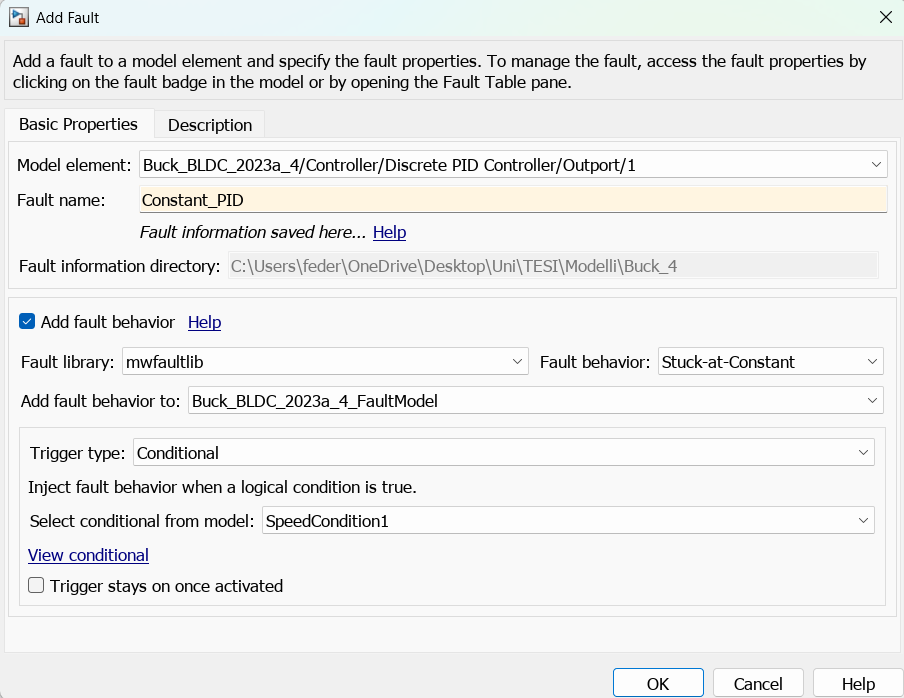}
    \caption{Adding new faults with the \SimulinkFaultAnalyzer.}
    \label{fig:process}
\end{figure}

To insert a fault, engineers should select the signal (or component) where the fault needs to be applied and use the graphical interface presented in \Cref{fig:process}.
The interface displays the model element related to the fault (``\emph{Model element}'') and enables engineers to set the fault name (``\emph{Fault name}'') as well as to define its behavior (``\emph{Add fault behavior}'').
\SimulinkFaultAnalyzer maintains the fault logic encapsulated in specific blocks.
This choice forces the fault logic to be separated from the model logic, thus preventing it from inadvertently becoming part of the model. 
To define the fault behaviors, engineers can reuse existing faults from a predefined library (``\emph{Fault library}'').
\Cref{tab:faultmodels} presents the list of faults provided by the \SimulinkFaultAnalyzer (``\emph{Fault behavior}'') and a short description (``\emph{Description}'').
For example, for the ``\emph{Constant\_PID}'' from \Cref{fig:process}, the engineer selected the ``\emph{Stuck-at-Constant}'' fault behavior from the library \emph{mwfaultlib}, thereby forcing the signal to assume a constant value.

\begin{table}[t]
\centering
\caption{Fault behavior}
\label{tab:faultmodels}
\begin{tabular}{l p{5.8cm} }
\toprule
\textbf{Fault behavior} & \textbf{Description} \\ \midrule
Add Noise & Adds random noise to the signal.\\
Negate Value & Negates the value of a signal.\\
Absolute Value & Transforms the signal into its absolute value.\\
Stuck-at-Ground & Sets the signal to the ground.\\
Gain & Adds a gain to the signal.\\
Offset-by-1 & Adds an offset to the input signal.\\
Unit Delay & Delays the signal by a single time unit.\\
Stuck-at-Constant & Sets the signal to a constant value. \\ \bottomrule
\end{tabular}
\end{table}

\begin{table}[t]
\centering
\caption{Trigger types}
\label{tab:triggers}
\begin{tabular}{l p{5.8cm} }
\toprule
\textbf{Trigger type} & \textbf{Description} \\ \midrule
Conditional & Activates a fault when a specific condition, on any of the signals or measures available in the model, is satisfied.\\
Timed & Activates a fault after the specified simulation time.\\
Manual & Activates a fault manually during the simulation by clicking on a button.\\ \bottomrule
\end{tabular}
\end{table}

Further, faults can be activated via triggers.
\Cref{tab:triggers} presents the list of triggers provided by the \SimulinkFaultAnalyzer (``\emph{Trigger type}'') and a short description (``\emph{Description}'').
For example, for the ``\emph{Sector\_fault}'' from  \Cref{fig:process}, the engineer selected the ``\emph{Conditional}'' trigger, 
and specifies a condition (``\emph{SpeedCondition1}'') that forces the fault to be activated when the speed becomes greater than $50 \; rpm$.

\begin{figure}[t]
    \centering   \includegraphics[width=.96\columnwidth]{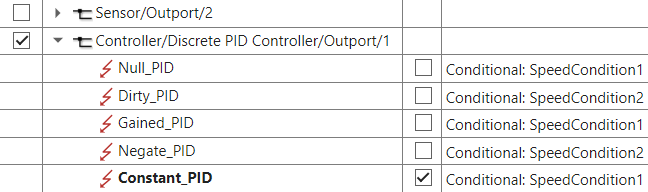}
    \caption{A portion of the Fault Table for our e-Bike.}
    \label{fig:table}
\end{figure}

\begin{figure}[t]
    \centering   \includegraphics[width=.96\columnwidth,clip, trim={0cm 0.05cm 0cm 0.28cm}]{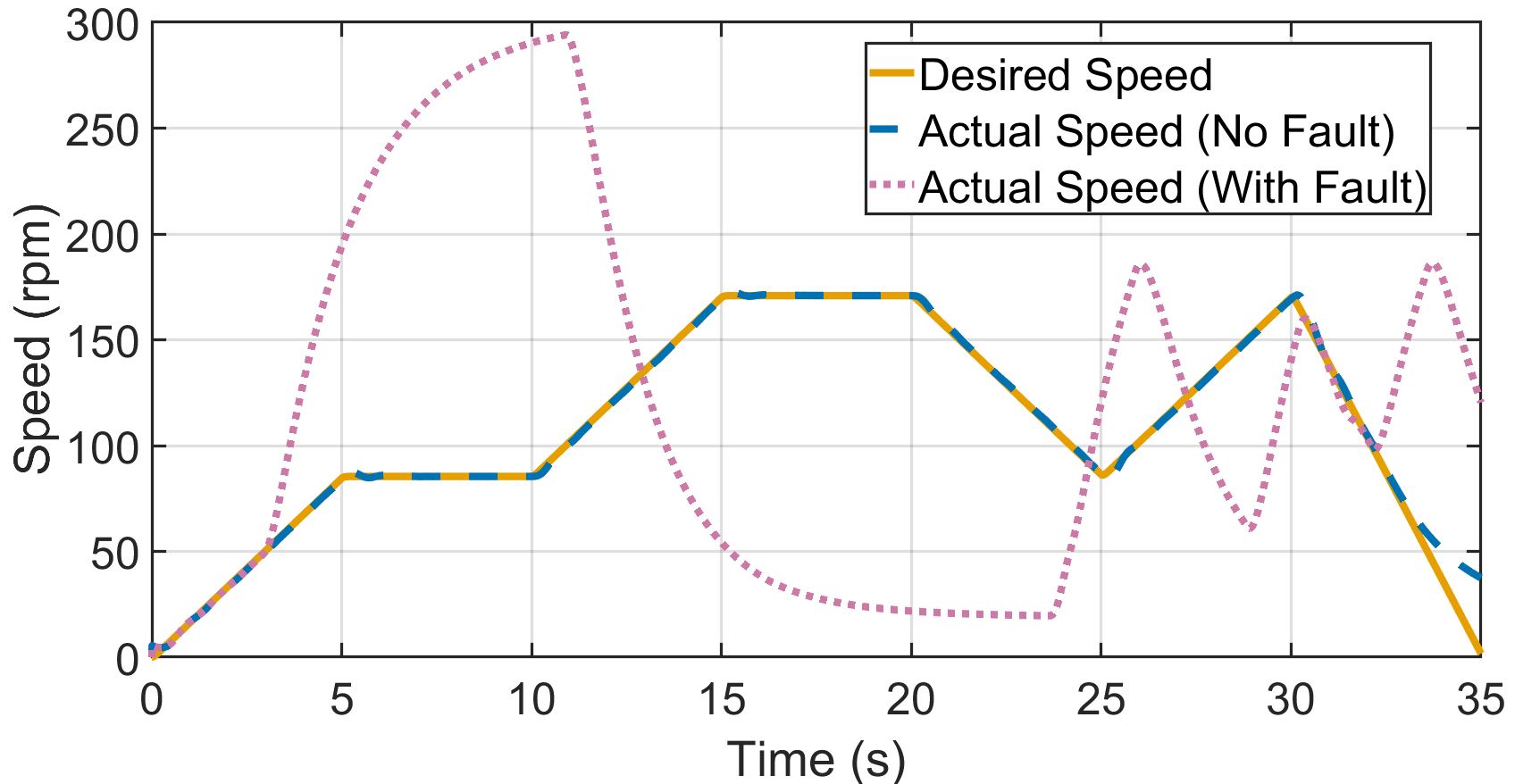}
    \caption{The desired speed and the speed of the e-Bike when the fault is activated and deactivated.}
    \label{fig:effect}
\end{figure}

\subsection{Support for Effect Analysis - EA}
\label{sec:effects}
Simulation-driven FMEA supports EA by providing engineers the possibility of simulating the model with the faults injected to analyze their effects.

The \SimulinkFaultAnalyzer provides Fault Tables that enable engineers to activate different faults.
For example, \Cref{fig:table} presents (a portion of) the Faults Table for our e-Bike model. 
The engineers activated the fault ``\emph{Constant\_PID}'' defined in \Cref{fig:process} and disabled all the other faults.

Then, simulating the model within \Simulink enables the analysis of the fault effect.
For example, \Cref{fig:effect} depicts the desired speed (yellow line), the speed of a simulation where the fault is not enabled (blue line), and the speed of the current simulation (pink line).
The simulation reveals that when the condition on the speed (``\emph{SpeedCondition1}'') is triggered --- i.e., when the motor speed exceeds $50 \; rpm$ --- the speed of the e-Bike suddenly stops following the desired speed.
It then increases until $300 \; rpm$, decreases until $20 \; rpm$, and begins oscillating.

This result enables engineers to understand the consequences of the fault. 
Specifically, forcing the gain of the PID to a constant value results in a significant increase of the speed of the e-Bike that exceeds the maximum speed of 25$km/h$ (i.e., 170 $rpm$ wheel speed, considering a 28-inch wheel), which is mandated by most European countries~\cite{Schepers2018safety,Schleinitz2017}.

Information regarding the effects of the faults helps engineers improve their models.
For example, engineers can decide to mitigate the fault ``\emph{Constant\_PID}'' by applying appropriate strategies (e.g., adding another PID that takes control when problems occur or adding saturation mechanisms).

\section{Evaluation}
\label{sec:evaluation}

To assess the usefulness of the Simulation-driven FMEA, we consider the following research questions (RQ):
\begin{itemize}
    \item \textbf{RQ1}: How helpful is the support from the \SimulinkFaultAnalyzer in \emph{modeling} the faults? (\Cref{sec:rq2})
    \item \textbf{RQ2}: How helpful is the Simulation-driven FMEA for \emph{safety} analysis? (\Cref{sec:rq3})
\end{itemize}

First, we present our benchmark (\Cref{sec:experimentalSetup}). 
Then, we answer RQ1 (\Cref{sec:rq2}) and RQ2 (\Cref{sec:rq3}).

\subsection{Benchmark Design} \label{sec:experimentalSetup}
To answer our research questions, two of the authors analyzed the e-Bike study subject and identified 15 faults (we later describe why our evaluation only considers 13 of these faults).
\Cref{tab:RQ1} provides the identifier (column ``ID'') and a textual description (column ``Description'') for each of these faults.
Our faults involve both the cyber (\emph{Controller}) and the physical (\emph{Sensor}) parts of the CPS specified in the lower and upper parts of \Cref{tab:RQ1}, and are applied both to \Simulink and \Simscape components.
For example, fault F1 is related to the sensor part and concerns the breakage of the speed sensor, while fault F12 concerns the software controller and involves some noise on the output of the PID.

To ensure that our evaluation is based on realistic faults, we interacted with our expert to assess the criticality, plausibility, and frequency of each fault.
We considered realistic faults that are significant for at least one of these criteria: We did expect critical faults to be less plausible and frequent than other faults.
Conversely, we considered nonrealistic faults that are both unplausible, unfrequent, and noncritical.

We conducted a structured interview with the designer of the e-Bike controller to identify realistic faults.
The designer had to assign a score between one and five for plausibility, frequency, and criticality; moreover the designer was required to justify their decision.
The lowest bound (0) represents lower plausibility, frequency, or criticality, and the highest one (5) corresponds to highly plausible, frequent, or critical faults.
\Cref{tab:RQ1} reports the score provided by the expert for the plausibility (Plaus.), frequency (Freq.), and criticality (Critc.) of each fault.
We considered a fault plausible, frequent, and critical if the score assigned by the expert is greater than or equal to $3$.

The majority of our faults are plausible ($66 \%$), critical ($74 \%$), and not frequent  ($74\%$).
Note that we expected the majority of faults to be infrequent, as undesirable behavior is not expected to be common.

We excluded faults F3 and F8 (identified with a blue background in \Cref{tab:RQ1}).
These faults involve an amplification error in the speed or sector sensors.
We considered them to be non-realistic as they are not plausible, critical, and frequent, i.e., they had both plausibility, frequency, and criticality values lower than $3$.
Therefore, our benchmark consists of the remaining $13$ realistic faults that are used to answer our RQs.


\begin{table}[t]
\centering
\caption{Plausibility, frequency, and criticality of the faults from our benchmark}
\label{tab:RQ1}
\setlength{\tabcolsep}{4pt}
\begin{tabular}{c l l l l}
\toprule
\textbf{ID}            & 
\textbf{Description}            & \textbf{Plaus.} & 
\textbf{Freq.} & 
\textbf{Critic.}  \\ \midrule
\textbf{F1} & Breakage of the speed sensor~\cite{TranSensorFaultDiagnosis}            & 4                     & 4                  & 5                    \\ 
\textbf{F2} & Noisy speed sensor~\cite{Murgano2021}                       & 4                     & 2                  & 1                    \\
\cellcolor{blue!25}\textbf{F3} &  \cellcolor{blue!25}Amplification error in the speed sensor  & \cellcolor{blue!25}1                     & \cellcolor{blue!25}1                  & \cellcolor{blue!25}1                    \\
\textbf{F4} & Delayed speed sensor~\cite{Dialynas02012023}                     & 1                     & 1                  & 3                    \\
\textbf{F5} & Speed sensor with constant value         & 4                     & 4                  & 5                    \\
\textbf{F6} & Breakage of the sector sensor            & 4                     & 4                  & 5                    \\
\textbf{F7} & Noisy sector sensor                      & 4                     & 2                  & 1                    \\
\cellcolor{blue!25}\textbf{F8} & \cellcolor{blue!25}Amplification error in the sector sensor & \cellcolor{blue!25}1                     & \cellcolor{blue!25}1                  & \cellcolor{blue!25}1                    \\
\textbf{F9}  & Delayed sector sensor~\cite{Dialynas02012023}                  & 1                     & 1                  & 3                    \\
\textbf{F10} & Sector sensor with constant value      & 4                     & 4                  & 5                    \\
\midrule
\textbf{F11} & PID controller design error            & 3                     & 2                  & 5                    \\
\textbf{F12} & Noise on the output of the PID         & 4                     & 2                  & 4                    \\
\textbf{F13} & Amplification of the output of the PID & 4                     & 2                  & 3                    \\
\textbf{F14} & Inverted PID output                    & 1                     & 1                  & 5                    \\ 
\textbf{F15} & Constant output of the PID             & 3                     & 2                  & 5                    \\\bottomrule
\end{tabular}
\end{table}

\subsection{RQ1: Support of Simulink Fault Analyzer} \label{sec:rq2}
To assess how the \SimulinkFaultAnalyzer helps in modeling faults, we proceeded as follows. 
\begin{itemize}
    \item \emph{Step~1}. We considered each of the 13 faults and modeled them.
The goal of this step is to assess whether we could model all the faults of our benchmark using the modeling support provided by the \SimulinkFaultAnalyzer (see \Cref{sec:modeling}). 
However, our models can potentially be inaccurate and not adequately reflect the expected fault behavior.
To mitigate this problem, we performed an additional step.
\item \emph{Step~2}. We conducted a structured interview with our e-Bike expert to determine whether the models of our faults were appropriate.
For inaccurate models, we tried to fix them using the modeling support provided by the \SimulinkFaultAnalyzer.
\end{itemize}

\emph{Results}. \Cref{tab:RQ1_models} presents the models of the faults from \emph{Step~1} (\Cref{tab:RQ1}).
Note that, as we will clarify in the following account, for many faults, multiple models (determined by multiple trigger types) were produced.
\Cref{tab:RQ1_models} provides the identifier of the modeled fault (column ``ID''),  the modeled behavior (column ``Fault behavior''), and the selected trigger type (column ``Trigger type'').
For certain faults, we could not decide whether a ``Timed'' or a  ``Conditional'' trigger was more appropriate because we did not discuss the triggers when we identified our faults with our domain expert (\Cref{tab:RQ1}).
Therefore, we proposed alternative trigger options for discussion with our expert.
For example, for fault F1, we considered the behavior ``Stuck-at-Ground'' and two triggers: Timed and Conditional.


\begin{table}[t]
\centering
\caption{Identifier (ID), fault behavior, and trigger type for all the models of the faults of our benchmark}

\label{tab:RQ1_models}
\setlength{\tabcolsep}{4pt}
\begin{tabular}{c l l l}
\toprule
\textbf{ID} & \textbf{Fault behavior} & \textbf{Trigger type} \\ \midrule
F1   & Stuck-at-Ground & Timed \& Conditional \\ 
F2 & Add Noise & Timed \& Conditional \\ 
F4 & Unit Delay & Timed  \& Conditional \\ 
F5  & Stuck-at-Constant & Timed \& Conditional\\ 
F6 &  Stuck-at-Ground & Timed \&  Conditional\\ 
F7 & \cellcolor{blue!25}Add Noise & \cellcolor{blue!25}Timed \& Conditional \\ 
F9 & Unit Delay & Timed \& Conditional \\ 
F10   & \cellcolor{blue!25}Stuck-at-Constant & \cellcolor{blue!25}Timed \& Conditional \\ 
F11   & \cellcolor{blue!25}Stuck-at-Ground & \cellcolor{blue!25}Conditional \\ 
F12    & Add Noise & Conditional \\ 
F13  &  \cellcolor{blue!25}Gain & \cellcolor{blue!25}Conditional \\ 
F14  &  \cellcolor{blue!25}Negate Value & \cellcolor{blue!25}Conditional \\ 
F15  &  \cellcolor{blue!25}Stuck-at-Constant & \cellcolor{blue!25}Conditional \\ 
\bottomrule
\end{tabular}
\end{table}

The feedback from the domain expert (\emph{Step~2}) confirmed that $53.8 \%$ (7 out of 13) of the models selected to model the faults were accurate.
For the faults that had multiple trigger options (F1, F2, F4, F5, F6, F7, F9, and F10), the expert confirmed that ``Conditional'' was the appropriate trigger.

Further, the expert confirmed that $46.2 \%$ (6 out of 13) of our faults were correctly modeled, except for a few minor inaccuracies.
Faults with minor inaccuracies are reported in \Cref{tab:RQ1_models} with a blue background color. 
The minor inaccuracies have been fixed by considering the feedback provided by the domain expert.


%

\begin{table*}[t]
\centering
\caption{Expected effect from the expert}
\label{tab:RQ2expected}
\begin{tabular}{l p{16cm}}
\toprule
\textbf{ID} & \textbf{Expected Effect}  \\ \midrule
F1 & The PID will have a high duty cycle. Additionally, at least one of the three current phases (and the battery currents) will show high values. 
\\
F2 & The duty cycle will reach its maximum value. \\
F4 & The PID will take more time to regulate the speed. \\
F5 & The PID will increase the duty cycle, leading to an actual speed higher than the desired one. \\
F6 & The system will stop because it is unable to encode the value, which lies between 1 and 6. \\
F7 & The system will show extremely high currents, and the motor will stop spinning.  \\
F9 & The system will stop. \\
F10 & The system will show very high phase currents until the motor slows down and eventually stops, then remains stationary.\\
F11 & The duty cycle will fall to zero, and the motor will slow down until it stops. \\
F12 & Phase currents will significantly oscillate.  \\
F13 & The PID will saturate, and the system will try to accelerate even if unnecessary.  \\
F14 & The system will slow down until it stops. \\
F15 & The system will accelerate the motor regardless of speed, until electrical equilibrium is reached or until the fault disappears. \\ \bottomrule
\end{tabular}
\end{table*}

\begin{table*}[t]
\centering
\caption{Experimental effect from Simulation-driven FMEA}
\label{tab:RQ2experimental}
\begin{tabular}{l p{16cm}}
\toprule
\textbf{ID} & \textbf{Experimental Effect} \\ \midrule
F7 & \emph{The simulation stops with an error message since the value ``0'' for the sector signal is not included within the accepted range [1-6].}\\
F10  & The system shows very high phase currents. When the fault is triggered, the measured speed shows marked oscillations around zero, if the initial speed is low, or it decreases slowly, showing evident oscillations. \emph{When the fault is deactivated, the PID maximizes the duty cycle to restore the speed as desired from the desired speed input. This process is very aggressive, especially if compared to the scenario in which the fault is not triggered, where smoother control actions are performed.} \\
F12  & \emph{Constant, but slight,} oscillations in the phase currents. 
\\
F13  & \emph{The PID shows evident oscillations when the fault occurs}, then restores standard behavior. \emph{There is no relationship between the initial value of the PID and the experimental outcome.} \\
F15 & The system accelerates the motor as much as possible, with duty cycle 1, then drops to 0 when the fault deactivates. \emph{This causes an alternation of very high and very low measured speeds, and also of very high and very low phase currents, both when the fault alternates between active and inactive.}\\ \bottomrule
\end{tabular}
\end{table*}

To summarize, \SimulinkFaultAnalyzer enables us to model all the \numfaults realistic faults of our benchmark.
In this process, we did not identify any major limitations in the support provided by the \SimulinkFaultAnalyzer. Nevertheless, we learned a few lessons (see \Cref{sec:discussion}) that may help further improve this tool.

\begin{Answer}[RQ1: Tool support]
\SimulinkFaultAnalyzer successfully helped us design the models for the 13 faults of our benchmark.
\end{Answer}

\subsection{RQ2: Usefulness of FMEA for Safety Analysis} \label{sec:rq3}

To assess the usefulness of FMEA for safety analysis, we proceeded as follows. 
\begin{itemize}
    \item \emph{Step~1}. We collected feedback from the expert regarding the expected effects of the faults by conducting a structured interview. During the structured interview, we asked what the expected system behavior was when each of the faults from \Cref{sec:rq2} was activated and recorded their answer.
    \item \emph{Step~2}. We used the support for EA provided by Simulation-driven FMEA (\Cref{sec:effects}) to analyze the effect of the fault.
    \item \emph{Step~3}. We compared the expected effect from the expert and the experimental effect from Simulation-driven FMEA.
\end{itemize}

\emph{Results}.
For each fault model presented in \Cref{tab:RQ1_models}, \Cref{tab:RQ2expected} describes the expected effect from the expert (\emph{Step~1}).
\Cref{tab:RQ2experimental} reports faults for which the experimental results did not match the expectations of the engineer (\emph{Step~2} and \emph{Step~3}). 
The portion of the effect that did not match the expectation is highlighted in italic. 
The expectation of the engineer is confirmed in $61.6\%$ ($8$ out of $13$) of the faults. 
For the remaining $38.4\%$ ($5$ out of $13$) of the faults, the simulation revealed effects that deviated from the engineer's expectation. 

We observed that the differences identified during our experiments deviated from experts' expectations for two reasons. 
In certain cases (i.e., F7), the expected outcomes did not align with the actual results obtained during simulation. 
For example, for fault F7, while the expert expected ``\emph{extremely high currents until the motor stops spinning},''  The Simulation stopped with an error message caused by the sector signal having a value of ``0'', which falls outside the accepted range of 1 to 6. 
This error is due to a bug in the modeling of the three-phase inverter.
The engineer did not consider the case in which the component could receive invalid values.
Thus, the three-phase inverter does not correctly handle invalid sector values, thus causing the simulation to stop prematurely instead of exhibiting the expected system behavior in the presence of the fault. Ideally, the system should perform an emergency stop and slowly decelerate till the motor completely stops. 
In other instances (i.e., F10, F12, F13, and F15), while the outcomes partially matched the expectations, additional effects were observed during the experiments, or minor inaccuracies emerged.
For example, for fault F13, the expert expected the PID to force the system to accelerate, even if this was unnecessary. 
The simulation revealed that the PID output had significant oscillations when the fault occurred, which was not expected.
Therefore, this analysis demonstrates the effectiveness of the Simulation-driven FMEA supported by \SimulinkFaultAnalyzer. 

\begin{Answer}[RQ2: Usefulness]
FMEA was useful for analyzing the safety of our e-Bike. 
It enabled the engineers to uncover fault side-effects (F10, F12, F13, and F14) that had previously been overlooked. In addition, it also helped identify unexpected effects~(F7).
\end{Answer}
\section{Discussion and Lessons Learned}
\label{sec:discussion}
We present the lessons learned (\Cref{sec:lessons}), discuss them (\Cref{sec:discussion_ss}), 
and summarize threats to the validity of our findings and experiments (\Cref{sec:threats}).

\subsection{Lessons learned} 
\label{sec:lessons}
We held joint meetings with MathWorks experts to describe and analyze our results and outline the lessons learned from this study.
The feedback from MathWorks experts was pivotal to confirm a proper usage of the \SimulinkFaultAnalyzer, its modeling capabilities, and the Simulation-driven FMEA support.
In the following account, we outline our lessons learned by classifying them into two categories: \Simulink Specific and \Simulink Agnostic.\\

\emph{\Simulink-specific}. These lessons learned are related to the usage of the \Simulink environment. They are useful for software engineers working as \Simulink engineers and for users of the \SimulinkFaultAnalyzer.

\begin{itemize}
    \item \emph{Lesson~1 (Triggers)}. 
    \SimulinkFaultAnalyzer offers several types of triggers, including manually activated, activated at a specific time instant (timed), or activated when a specific condition is true (conditional). 
    For our e-Bike example, we had to model an event-based trigger.  
    For example, the trigger for fault F1 (Stuck-at-ground) applies when the sensor breaks. 
    This problem is likely to occur when the desired speed undergoes a sudden change --- for example, in the case of a sudden acceleration or braking.
    Therefore, the fault should apply to the measured speed when the sensor suddenly breaks and remains broken till the end of the simulation.
    By default, conditional triggers activate the faults each time a specific system condition is true and stop the fault injection as soon as it becomes false.
    To force the fault to remain active until the end of the simulation, the flag ``Trigger stays on once activated'' must be activated (see \Cref{fig:event_based}).
    We initially overlooked this flag and defined the logic to detect the sudden change and to maintain the condition of the trigger to true until the end of the simulation within the \Simulink model (partially defeating the benefits of using the \SimulinkFaultAnalyzer). 
    The meeting with the \Simulink experts evidenced the presence of the flag, as depicted in \Cref{fig:event_based}. 
    We modified the model of the fault to appropriately utilize the features provided by the \SimulinkFaultAnalyzer. 

    \begin{figure}[t]
    \centering   \includegraphics[width=.8\columnwidth, clip, trim={0cm 0cm 0cm 0.2cm}]{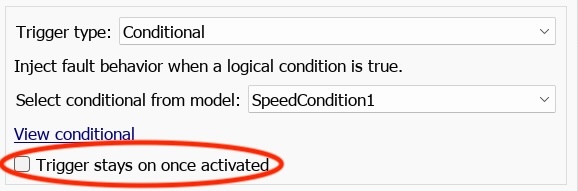}
    \caption{Flag enabling the use of event-based triggers.}
    \label{fig:event_based}
    \end{figure}

    \item \emph{Lesson~2 (Target of the fault application)}. 
   For \Simulink models, faults are typically applied to physical signals, except for the \Simscape portion of the models~\cite{FaultAnalyzerCompatibleBlocks}.
   For \Simscape, faults are applied to components rather than to signals.
   We believe that extending \Simscape by providing the opportunity to add faults to signals can benefit practical applications.
    For example, fault F2 requires adding noise to the speed sensor. 
    This fault may occur when electromagnetic noise is present on the controller board's cables.
    However, to directly apply the fault to an electrical (\Simscape) signal, it is necessary to drive a \Simscape component (e.g., a variable resistor~\cite{VariableResistor})
    that changes the resistance depending on the value of a \Simulink signal that is affected by the fault. 
    \item \emph{Lesson~3 (Single fault activation on a specific signal)}. 
    \SimulinkFaultAnalyzer enables the activation of multiple faults (related to different signals) within a single simulation using the multiple simulations panel~\cite{MultipleFaults}.
    However, for example, activating both F1 and F2 requires simultaneously activating two faults on the same signal. 
    This situation either requires (a) inserting the two faults at the source and the destination of the connection, respectively, or (b) creating an additional fault behavior that mimics the scenario in which both faults are activated.  
    Note that while the second solution also applies to situations in which more than two faults need to be activated, the first solution only applies to the case of two faults.
    Extending the \SimulinkFaultAnalyzer by enabling the activation of multiple faults on the same signal will avoid the creation of additional fault behavior.
    \item \emph{Lesson 4 (Activation of a fault on multiple signals)}. 
    In certain cases, we had to simultaneously apply the same fault to multiple signals.
    For example, fault F7 requires considering noise on the sector sensor caused by electromagnetic interference.
    In this case, the same noise should be applied to more than one signal, i.e., to all cables in the same area of the controller's board.
    To model this situation, we modified the model by adding a \Simulink block that generated a constant signal with value ``0'', adding the noise fault to this signal, and summing the value of this signal to all signals subjected to electromagnetic interference.
    In this manner, when the fault is activated, the noise is applied to all the signals subjected to electromagnetic interference.
    Extending the \SimulinkFaultAnalyzer by providing the opportunity to apply the same fault to multiple signals would avoid modifying the model under analysis.
    \item \emph{Lesson 5 (Accelerator Execution Mode)}. Unlike the normal execution mode, the accelerator mode uses Just-in-Time (JIT) acceleration to generate an execution engine in memory instead of generating C code or MEX files~\cite{Acceleration}. This mode increases performance and reduces the simulation time. 
    The \SimulinkFaultAnalyzer does not currently support the accelerator mode.
    This can be a useful addition to support quick prototyping. 
    In our scenario, the accelerator mode would have enabled us to save approximately 30 seconds for each simulation.
\end{itemize}


\vspace{0.2cm}
\emph{\Simulink-agnostic}. These lessons are general and not specific to the Simulink environment. They are valuable for safety analysts, regardless of the framework they use.

\begin{itemize}
    \item \emph{Lesson 6 (Fault modeling --- \Cref{sec:rq2})}. Explicitly modeling faults enabled us to discuss and reason with our e-Bike expert regarding the possible causes for the system's malfunctions. 
    Our e-Bike expert confirmed that this practice helped with reasoning regarding the causes of possible safety breaches. 
    These faults were not explicitly identified, discussed, and documented earlier.
    It also enabled the analysis of conditions (modeled by triggers) that activate these faults.
    This enabled the strengthening of the reasoning regarding situations and conditions that are more ``stressful'' for the e-Bike software and its safety, and thus, more likely leading to faults.
    It encouraged deeper reflection, which led to the identification of faults that were not considered earlier.
    For example, the engineer identified additional faults, such as those related to the battery degradation or those involving the inverter, that they had never considered in their design.
    These faults are added by the engineer pipeline to continuously assess the safety of the e-Bike.
    To summarize, explicitly modeling faults can benefit CPS applications as they support a more rigorous safety analysis. 
    \item \emph{Lesson 7 (Fault simulation --- \Cref{sec:rq3})}. Engineers typically assess the safety of their system by manually defining inputs and visually inspecting the models' outputs. This is a common practice in industrial environments, particularly when preliminary and high-level models of the system (such as the one we considered) are evaluated.
    However, this practice may lead engineers to overlook certain possible issues and the effects of these faults.
    Having a framework that enables the automatic analysis of CPS failure, such as the \SimulinkFaultAnalyzer, mitigates this problem.
    The engineer who developed the model utilized in this study acknowledged that the proposed Simulation-driven FMEA approach was significantly beneficial for reasoning about plausible faults and their effects. 
    Using this framework enables the identification of discrepancies between the expected and observed effects for certain faults (\Cref{sec:rq3}). 
    For example,    \Cref{fig:effect} depicts the measured speed of the e-Bike when fault F15 occurs, thereby highlighting the maximum speed reached (in orange) and the regions where significant deceleration is observed (in purple).
    First, it can be noted that when F15 is active, the e-Bike reaches a maximum speed of approximately $40$ km/h, which is above the maximum legal speed limit ($25$ km/h)~\cite{EURegulation} for an e-Bike. 
    This limit is exceeded when the fault is active.
    Moreover, the sections highlighted in purple in \Cref{fig:effectLessons} exhibit a deceleration of approximately $2$ m/s$^2$ --- i.e., about $0.20$ g --- which is within the same order of magnitude as the deceleration of a commercial aircraft during a normal landing — typically between 1.54 $m/s^2$ and 3.09$m/s^2$ — and approximately one-fifth of the deceleration experienced during an emergency stop (rejected take-off), which can reach 5.14 $m/s^2$~\cite{DecellarationAirplane}.
    However, the critical issue is that such braking is neither expected nor commanded by the user, which poses safety and stability risks for both the vehicle and the rider.
    To summarize, fault simulation support benefits CPS designers in performing a more systematic and rigorous analysis of the effects of faults. 

 \begin{figure}[t]
    \centering   \includegraphics[width=.96\columnwidth]{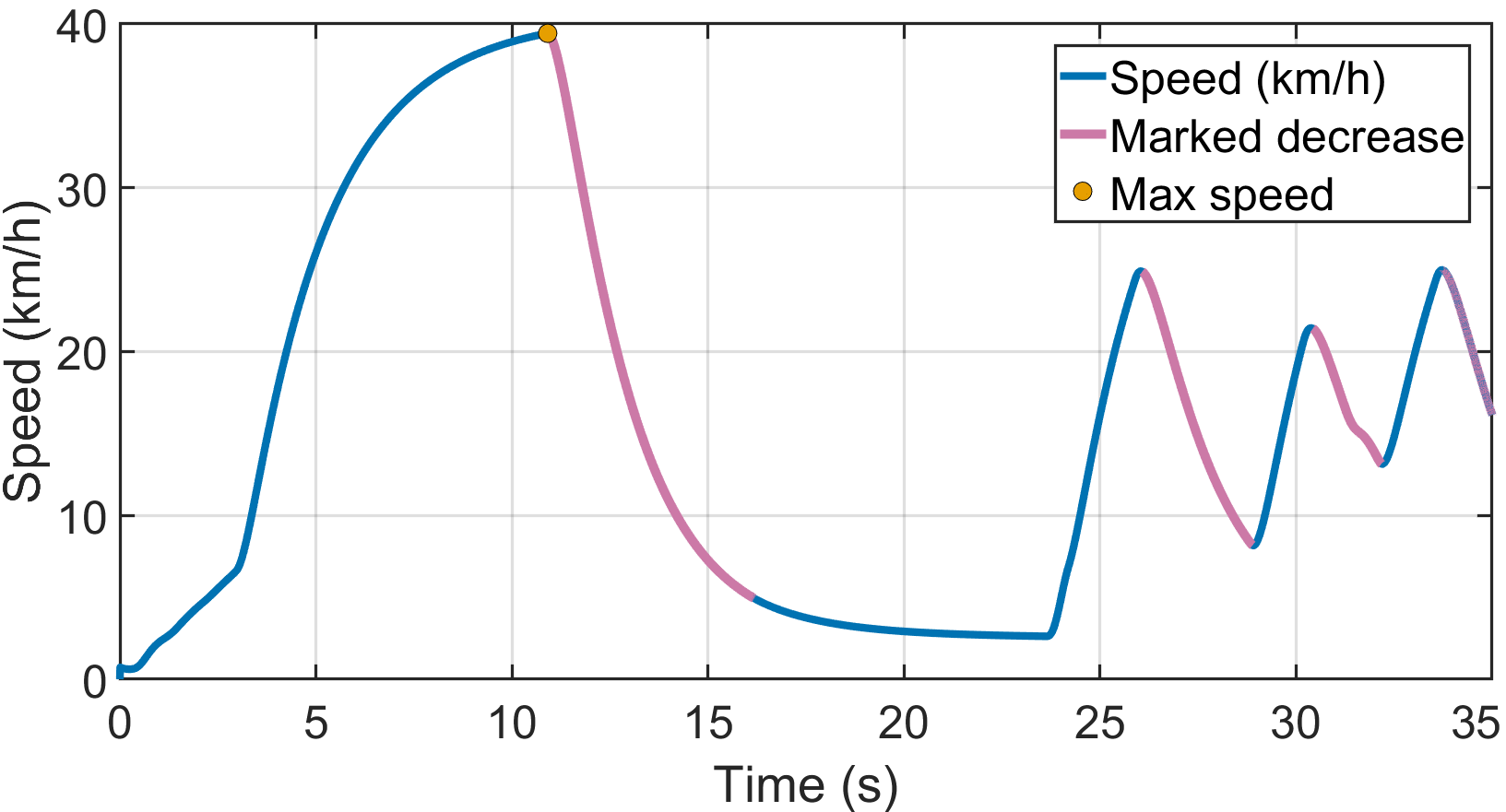}
    \caption{The measured speed of the e-Bike when fault F15 is activated, thereby highlighting the peak speed and the marked decrease intervals.}
    \label{fig:effectLessons}
    \end{figure}
    \item \emph{Lesson 8 (Development of mitigations)}. 
    Our expert not only modeled and analyzed the effects of faults but also initiated reasoning regarding possible mitigations.
    During our informal discussions preceding the structured interviews, the expert provided possible mitigations that could have been activated when specific faults occurred. 
    For example, for fault F2 (noisy speed sensor), adding a backup sensor is a possible mitigation.
    The control logic switches the active sensor when it detects that the noise on the monitored speed exceeds a certain threshold.   
    \Cref{fig:mitigation} depicts the system behavior when this mitigation is activated. 
    \Cref{fig:mitigation:a} presents the desired speed and the monitored speed by the original (and faulty) sensor, while \Cref{fig:mitigation:b} illustrates the measured speed when the mitigation is applied.
    \Cref{fig:mitigation:c} depicts the activation status of the mitigation strategy: When mitigation is \emph{On}, the backup sensor is used; when it is \emph{Off}, the speed is read from the original sensor.
    The mitigation enables the system to switch from sensor~1 to sensor~2 when the fault is activated. 
    This mitigation strategy enables the e-Bike to follow the desired speed and reduce the fault effect. 

    \begin{figure}[tb]
        \centering
        \begin{subfigure}[t]{\linewidth}
            \centering
            \includegraphics[width=.96\linewidth,clip,trim={0cm 15cm 0.8cm 0cm}]{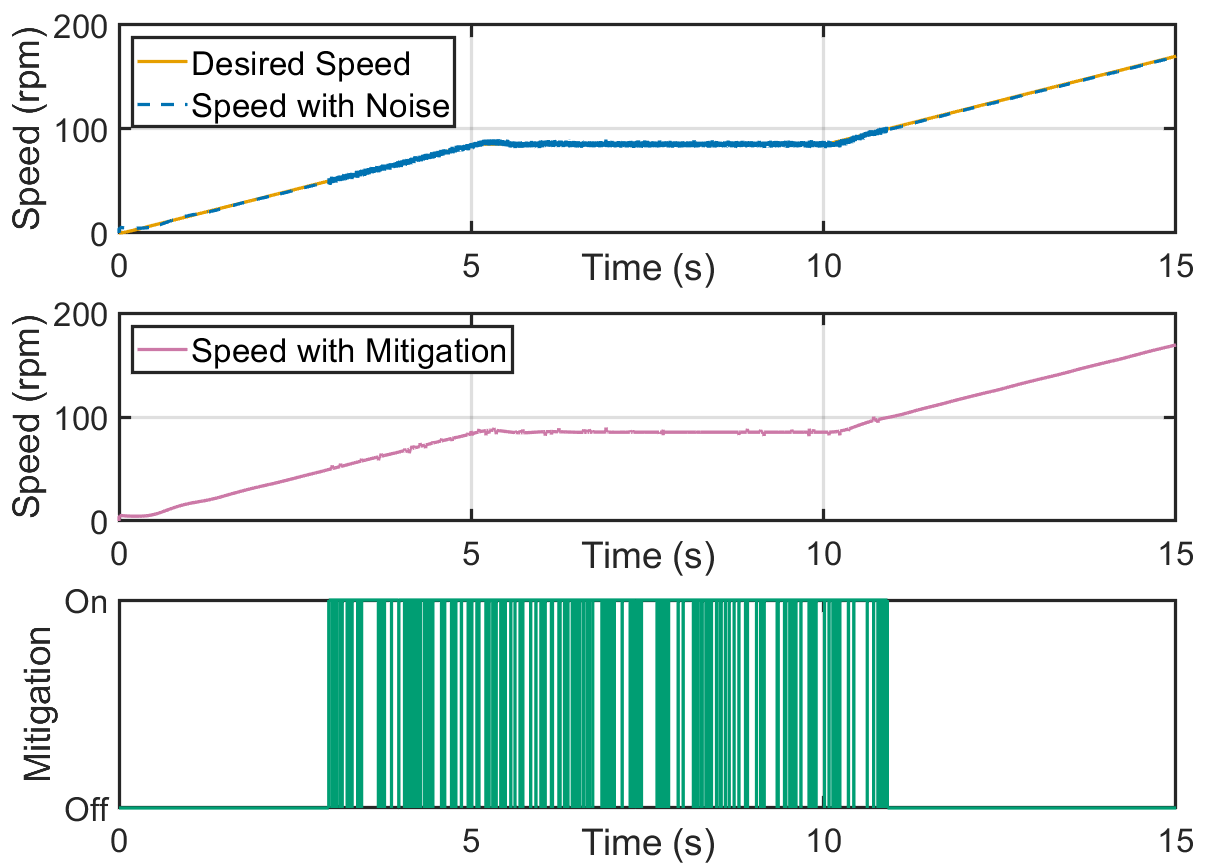}
            \caption{\label{fig:mitigation:a}The desired speed and measured speed of the e-Bike without mitigation.}
        \end{subfigure}
        
        \medskip
        
        \begin{subfigure}[t]{\linewidth}
            \centering
            \includegraphics[width=.96\linewidth,clip,trim={0cm 7.5cm 0.8cm 7.9cm}]{figures/Mitigation.png}
            \caption{\label{fig:mitigation:b}The measured speed of the e-Bike with mitigation.}
        \end{subfigure}
        
        \medskip
        
        \begin{subfigure}[t]{\linewidth}
            \centering
            \includegraphics[width=.96\linewidth,clip,trim={0cm 0cm 0.8cm 15.5cm}]{figures/Mitigation.png}
            \caption{\label{fig:mitigation:c}Mitigation activation status.}
        \end{subfigure}
        \caption{Mitigation of fault F2. Two sensors are used to measure the speed, and the backup sensor is activated when the level of noise surpasses the defined threshold.}
        \label{fig:mitigation}
    \end{figure}
    
    \item \emph{Lesson 9 (Mindset instauration)}. Nowadays, developing safer and more secure systems is becoming pivotal, particularly considering that new CPS are ubiquitous, more connected, and perform more critical activities. 
    Therefore, it is necessary to instantiate a proper engineering mindset based on rigorous and systematic safety reasoning.
    The results of this study confirm that a proper interaction among software engineers, system experts, academics, and industries developing tools for safety reasoning can help create this mindset by sharing knowledge, experiences, and solutions.
    \item \emph{Lesson 10 (Software engineering for e-Bikes)}. Bikes are traditionally not controlled and regulated by software.
    Therefore, e-Bike design typically considers both electrical aspects (e.g., the performance of the e-Bike) and physical aspects related to traditional bikes (e.g., material selection and aerodynamics).
    This traditional view is also reflected in recent analyses of e-Bike safety studies (e.g.,~\cite{cherry2019bike}) that mostly consider the behavior of the rider, maximum speed of the electrical motors, etc.
    However, as software becomes a central part of these vehicles, safety analysis cannot be limited to hardware. It must also consider software-related aspects that may lead the vehicle to violate the safety standards and behave improperly. 
    The results of this study reveal that as e-Bikes are becoming software systems, software engineering techniques are needed to support rigorous development as well as the safety analysis of these systems.
\end{itemize}


\subsection{Discussion} \label{sec:discussion_ss}
In this section, we discuss the originality of our work, the relevance of our findings for industrial applications,  the significance of our contributions, and the generalizability of our results.

In this paper, we applied the Simulation-driven FMEA approach to a case study in the e-Bike domain --- the e-Bike motor controller component.
This contribution is \emph{original}: Simulation-driven FMEA is a recent technology~\cite{rhein2024simulation,SimulinkFaultAnalyzer} that has never been applied to the e-Bike domain.
Therefore, it improves the current state of the practice in the e-Bike industrial context.

In addition, our results are \emph{relevant} to the e-Bike industry, as e-Bikes are software-intensive systems~\cite{SofwareBike}. 
These systems are safety-critical and must comply with safety regulations and requirements~\cite{Schepers2018safety,Schleinitz2017}. Our results can also be potentially generalizable to other industrial domains (see \Cref{sec:lessons}).

Further, this work provided \emph{significant} contributions. 
Compared with previous work assessing Simulation-driven FMEA on a robotic~\cite{miragliaRobot} and avionic~\cite{rhein2024simulation} case studies, this work provides significant \Simulink-specific and \Simulink-agnostic lessons learned that can support the usage of Simulation-driven FMEA in practical scenarios. 
These lessons learned have been considered valuable by \Simulink experts.

Finally, our work provides significant \emph{generalizable} lessons learned.
We discussed how our results on fault modeling and simulation apply to contexts beyond the specific industrial context (e-Bikes) of our paper.

\subsection{Threats to Validity} \label{sec:threats}
The fault models and triggers we designed threaten the \emph{internal validity} of our results; other models and triggers could lead to different results. 
However, the fact that all models and triggers provided by the \SimulinkFaultAnalyzer have been used at least once mitigates this threat because it ensures that each model and trigger has been considered by our evaluation.

We do not claim that \emph{all} our results can necessarily be generalized to study subjects from other domains. 
We acknowledge that the selection of our study subject (a model from the e-Bike domain) and the faults could threaten the \emph{external validity} of our results, as it affects their generalizability.
However, the fact that we validated them through an interview with the engineer who developed the model mitigates this threat since it ensures that the faults are representative (at least) for this domain.
Furthermore, our lessons learned result from meetings with \Simulink experts who possess a comprehensive understanding of many CPS domains and needs.

The process we applied in our study is general and can be potentially applied to other study subjects.
Future studies will assess if and how our results generalize to other domains.
They can confirm or refute our hypotheses for other types of systems.

\section{Related Work}
\label{sec:related}
This section (a)~summarizes other techniques that support engineers in analyzing the safety of their systems, briefly justifies the choice of Simulation-driven FMEA, and (b)~presents other works that analyze the software of e-Bikes.

\vspace{.5em}
\emph{Support for safety analysis}. FMEA~\cite{FMEA} is a widely used reliability analysis tool for identifying and mitigating failures in systems, designs, and processes~\cite{8234630}. 
Simulation-driven FMEA~\cite{rhein2024simulation} integrates the simulation capabilities of existing tools within FMEA.
This technique was initially demonstrated by considering a flight control system of an unmanned ultralight helicopter~\cite{rhein2024simulation}. 
Unlike the original work~\cite{rhein2024simulation}, which focuses on presenting the proposed solution, this paper empirically evaluates the effectiveness of Simulation-driven FMEA in an e-Bike case study to provide practitioners with guidelines and lessons learned.

There are many techniques and approaches similar to FMEA that support engineers in risk evaluation and reliability checks. 
For example, data-driven FMEA~\cite{Ervural2023A} relies on historical data and system metrics to evaluate and prioritize risks, life cost-based FMEA~\cite{RHEE2003179} measures risks for the system life cycle costs, model-based safety analysis~\cite{Joshi2006} extends a shared system model with faults and physical elements to enable automation, and neuro-fuzzy techniques~\cite{Ivančan2023Improvement} manage the uncertainty and the subjectivity of risk evaluations. 
Other studies integrate Monte Carlo simulations to manage uncertainty in complex scenarios, integrate FMEA with a widespread reliability and safety model technique (GO methodology)~\cite{Liu2019Enhanced}, and propose a multiperspective FMEA method based on rough number projection~\cite{AN2025109192} 
to obtain more accurate and coherent safety evaluations~\cite{AN2025109192}. 
There are also solutions (e.g.,~\cite{10.1145/3650212.3680331,formica2023simulation,10.1145/3611643.3613894}) that automatically identify failure causes using testing techniques.
In our study, we decided to assess Simulation-driven FMEA since it automates the fault analysis and identifies its effects on the system and also because it is integrated with industrial solutions (\SimulinkFaultAnalyzer). 
In addition, recent studies~\cite{cherry2019bike,thieme2020incorporating} advocate a more extensive use of FMEA to analyze software failures, as safety analysis typically overlooks software failure because system designers usually (erroneously) assume that the system's software does not fail.
Our study considers both the cyber (software) and physical failures of the CPS (respectively specified in the lower and upper parts of \Cref{tab:RQ1}).

\vspace{.5em}
\emph{E-Bike software analysis.} 
Software plays a critical role in modern e-Bikes. 
A recent article~\cite{EBikeSoftware} clarifies that while most riders ``\emph{think the e-Bike difference has to do with the frame, motor, or battery, it’s usually all in the code}.''

Developing software for eBikes is costly~\cite{EBikeSoftware}.
To design their software, engineers need to have a holistic view of the system, including its motor and the battery.
Acquiring this knowledge is complicated and time-consuming. 
We experienced these factors in our project and had to invest significant time learning the know-how necessary to work within this domain.
Specifically, we significantly interacted with our domain expert to obtain the domain knowledge necessary to operate in this field.

The e-Bike market consists of a large number of players (over 10,000 companies offer e-Bikes and a 27.15 USD billion market in 2022~\cite{E-BikeMarket}) that typically develop proprietary software that is not publicly available.
Nevertheless, there are several open-source e-bike projects in existence.
For example, the OpenSource e-Bike project~\cite{OpenEBike} provides software to help users build their e-Bikes.

Recent studies confirm that recent changes in electric vehicles have introduced new security threats and additional software safety concerns~\cite{10.1145/3650210}. 
Standards have been developed to regulate the development of e-Bike software systems and ensure they are sufficiently safe. 
The ISO 26262 standard (Part 6)~\cite{ISO26262} specifies software safety requirements for road vehicles (including e-Bikes).
EN 15194~\cite{EN15194} is a standard dedicated to e-Bikes; for example, it specifies a maximum delay for the motor to react to user actions, such as starting or stopping pedaling.
Further, safety breaches were found in a few e-Bikes. In 2020, VanMoof updated their app since customers could increase their pedal-assisted power beyond the EU limit of 25 km/h by switching to the US country setting in their app: in the US, e-Bikes are generally capped at 32 km/h (20 mph)~\cite{SofwareBikeCheating}.

Despite these safety concerns, limited research has considered software engineering techniques for e-Bikes.
A recent study considers the problem of automatically generating test cases for  e-Bikes~\cite{marzella2025test}.
It assessed the effectiveness of existing test case generation solutions (HECATE~\cite{formica2023simulation}) in identifying bugs on Simulink models from the e-Bike domain.
Our work is significantly different in its scope and objective (supporting the safety analysis of the system), technological support (\SimulinkFaultAnalyzer), and evaluation methodology. 
\section{Conclusions and Future Work}
\label{sec:conclusions}

In this paper, we applied Simulation-driven FMEA on an industrial case study in the e-Bike domain.
We used \SimulinkFaultAnalyzer as a tool to evaluate FMEA. 
Our objective was to assess the effectiveness and practicality of Simulation-driven FMEA for assessing and improving software safety in CPSs. 
We identified 13 realistic faults involving both cyber and physical parts, modeled them using \Simulink fault injection features, and evaluated their effects through simulation. 
Our results, which we evaluated with a domain expert and \Simulink experts, show that \SimulinkFaultAnalyzer effectively supports fault modeling and effect analysis. 
For $38\%$ of the cases, simulation results deviated from the engineer’s initial expectations, thereby revealing unexpected fault impacts, enabling meaningful safety insights and design refinements, and allowing deeper engineering understanding and model improvement.
As a result of this experience, we identified a set of lessons learned, which involved both \Simulink-specific and \Simulink-agnostic considerations, and we presented them to \Simulink experts.
Our findings are useful for software engineers who work as \Simulink engineers, use the \SimulinkFaultAnalyzer, or work as safety analysts.

In future work, we plan to extend our study to other CPSs to investigate the generalization of our conclusion. 
We also plan to extend the Simulation-driven FMEA with the automatic classification of fault effects.

\section*{Data Availability}
The original \Simulink model, the model extended with the faults, the plots related to the results of the simulation, and the transcripts of our interviews are available at \cite{ReplicationPackage}. The replication package will be made publicly available and assigned a DOI upon acceptance.

\section*{Acknowledgments}
The authors thank Giovanni Miraglia, Mahesh Nanjundappa, and Kevin Barkett from The MathWorks Inc. for their support and valuable feedback on the \SimulinkFaultAnalyzer.
The work of Andrea Bombarda is supported by PNC - ANTHEM (AdvaNced Technologies for Human-centrEd Medicine) - Grant PNC0000003 – CUP: B53C22006700001.
The work of Marcello Minervini, Aurora Zanenga and Claudio Menghi was partially supported by the European Union -- Next Generation EU. ``Sustainable Mobility Center (Centro Nazionale per la Mobilit\'a Sostenibile - CNMS)'', M4C2 -- Investment 1.4, Project Code CN\_00000023.  
The work of Claudio Menghi was partially supported by projects SERICS (PE00000014) under the NRRP MUR program and GLACIATION (101070141).

\bibliographystyle{IEEEtran}
\bibliography{main}

\end{document}